\documentclass{article}

\usepackage{microtype}
\usepackage{graphicx}
\usepackage{subfigure}
\usepackage{booktabs} 
\usepackage{enumitem} 

\usepackage{hyperref}
\usepackage{xurl} 

\usepackage[accepted]{mlsys2025}

\mlsystitlerunning{Capital Efficiency, Isolation, Risk, and Ruin in Generative AI Inference Serving Infrastructure} 

\begin{document}

\twocolumn[
\mlsystitle{Capital Efficiency, Isolation, Risk, and Ruin \\ in Generative AI Inference Serving Infrastructure} 



\mlsyssetsymbol{equal}{*}

\begin{mlsysauthorlist}

\mlsysauthor{Ren{\'e}e St. Amant}{msft}
\mlsysauthor{Anjaly Parayil}{msft}
\mlsysauthor{Ankur Mallick}{msft}
\mlsysauthor{Victor R\"uhle}{msft}
\end{mlsysauthorlist}

\mlsysaffiliation{msft}{Microsoft} 
\mlsyscorrespondingauthor{Ren{\'e}e St. Amant}{reneestamant@microsoft.com} 

\mlsyskeywords{AI, Capital Efficiency, Isolation, Ergodic Ruin, Risk, MLSys, LLM, Inference Serving}

\vskip 0.3in
\begin{abstract}

GenAI serving is an extremely capital-intensive endeavor that demands rapid, step-change improvements in capital efficiency to prevent industry wide increases in the cost of capital from limiting growth plans. Keeping pace with the economic necessities will increasingly require infrastructure providers to confront operationally-inconvenient realities of production serving, including the extreme heterogeneity of GenAI request costs. We have explored workload isolation as a foundational, fleet-level lever for increasing capital efficiency. Isolation is the practice of intelligently grouping ``like" requests before execution. Rather than exposing every replica to the full heterogeneous workload distribution, isolation deliberately \emph{reshapes the workload presented to the hardware} by assigning requests with different cost characteristics to different portions of the fleet. Workload isolation on its own does not make expensive requests cheaper; it prevents a costly portion of the workload from unnecessarily determining the operating point and capital efficiency of the entire fleet. It additionally lays the foundational groundwork that can serve as a launching point for compounding gains via narrow-band optimizations (tailoring) and custom silicon designs while also informing up on pricing policy. This paper uses representative production data and experience to convey the motivation behind isolation as a fleet-level lever for increased capital efficiency.

\end{abstract}

]


\printAffiliationsAndNotice{}  


\section{Introduction: The Economic Context}
\label{sec:intro}

Generative AI has passed the threshold of technological novelty and entered a phase of ruthless economic reality. While foundation models have achieved unprecedented capabilities, the commercial viability of deploying them at scale hinges on a single, critical factor: capital efficiency.

\subsubsection{The Cost of Capital}
Capital efficiency in this context refers to the ability to monetize capital investments such that they become wealth producing assets in a way that is more efficient than the alternative capital deployments. Capital efficiency answers the question \emph{``How smart was a capital investment?"}

The infrastructure supporting GenAI workloads is deeply capital-intensive. The cost of GPU servers is multiples higher than CPU servers, even before accounting for facility upgrades supporting state-of-the-art cooling and power distribution. In 2026, the aggregate capital expenditure of the major hyperscalers including Amazon, Microsoft, Alphabet, Meta, and Oracle is projected to exceed a staggering \$700 billion, with approximately 75\% of this spend dedicated exclusively to AI infrastructure~\cite{kirsch2026bigtech}. To put this into historical context, this represents a near-doubling of capital deployment compared to prior years, forcing hyperscalers to raise unprecedented levels of debt or redirect free cash flow to fund data centers, AI silicon, and the power generation required to run them. With the technology sector approaching spending of a trillion dollars annually to lay the physical tracks for GenAI, we've passed an investor inflection point where the available risk capacity has potentially reached its limits and is at risk of contracting. 
The consequences for failing to provide step-wise gains in capital efficiency, which is monitizable goodput per dollar spent, is an increase in the cost of capital at best and a growth limitation from capital starvation at worst.  Additionally, to create positive economic advantage over time, it is not enough to improve return on invested capital (ROIC), you must do so at rate greater than the rate of change in the weighted average cost of capital (WACC).

\subsubsection{Additional Headwinds}
Further, we can expect to face increased demands for capital from certain structural headwinds within GenAI inference serving. 
Frontier models continue to grow in size requiring larger and larger GPU (or AI accelerator) footprints. Resource occupation times continue to grow with reasoning models, increased thinking depth, and long-horizon agentic workloads.
Per-user (and per-agent) usage intensity is increasing exponentially. The maximum context lengths supported by new model generations have rapidly changed in a short period of time. Finally, users increasingly expect not only their histories to be part of AI memory, but entire code repositories to be attached to reasoning requests, which further strains the capital requirements necessary to support the expected utility.

\subsubsection{Downward Pricing Pressure}
For any given level of intelligence, the per-token pricing is on an undeniable downward trajectory. Competitive pressures combined with a rapid cadence of GPU generations indicate a continuation of this trajectory, though specific bottlenecks may at some point find a floor due to resource scarcity. 
To maintain the same return on invested capital, costs must decrease at least as quickly, and a large percentage of the total cost is in the amortization of the AI hardware. 

\subsubsection{Supply Chain Constraints}
Compounding this financial pressure is the physical reality of the global supply chain. Compute and memory are not infinitely elastic resources, and organizations cannot simply buy their way to growth in the presence of physical supply chain constraints. Hardware scarcity imposes a hard cap on both top-line growth and the sheer volume of global requests that can be served. When silicon supply is finite, software and architectural efficiency becomes the only viable mechanism for expanding reach and scaling operations. For hyperscalers with multiple AI monetization strategies (frontier model development, inference serving, GPU hourly rentals, etc.) optionality exists for capital deployment, which inevitably leads to a competition for resources with \emph{relative capital efficiency} as the obvious hurdle for deciding how to deploy supply-constrained resources.

\subsubsection{Industrial Age Lens}
\label{sec:intro-industrial}
The economic viability of large-scale AI inference depends on how much useful work can be extracted from each provisioned capital investment, i.e. a GPU. To best understand this, we must shift our thinking from the highly-favorable economics of the information age (low-capital intensity) to an economic lens reminiscent of industrial manufacturing.  The infrastructure powering GenAI is not an abstract cloud resource; it is expensive machinery. In traditional manufacturing, capital-intensive machinery subject to steep depreciation cannot be allowed to run at low utilization. AI infrastructure faces the rapid cadence of silicon generations with architectural characteristics matched to the current frontier model architectures. Consequently, SaaS-era economics, forgiving of idle cycles and lower-than-capacity resource utilization, that prioritized an automatic connection between performance and growth \emph{must evolve to integrate capital efficiency as a first-order design constraint}.

Recognizing that utility is produced from output tokens (both internal and customer exposed), there is a financial imperative to maximize the \textit{lifetime output tokens} (utility) generated by these capital investments before the hardware becomes obsolete. 
Focusing on an output production unit is a particularly difficult challenge for computer scientists that have specialized within stack boundaries on processing input and ultimately necessitates an expansion of the frame to systems and cross-stack solutions that span previously isolated siloes. 

\subsubsection{Closing the Gap}
To bridge the gap between technological capability and business viability, the industry requires step-change increases in capital efficiency. Given the intersection of these economic forces with a tendency to treat efficiency as secondary to performance, we should expect the required pace of efficiency gains to become increasingly inconvenient to deliver. 
Achieving these step changes requires a holistic rethinking of how we serve requests and an integration of the material differences between GenAI serving and our previous SaaS context. 
Our work acknowledges that this economic reality requires the consideration of inconvenient paths forward.

\subsubsection{Isolation}
This paper describes the motivations behind the use of isolation as a specific fleet-level lever for improving capital efficiency.
We use representative data from production serving to demonstrate one important (and easy to understand) isolation mechanism: isolating requests by input sequence length.
Different workloads will expose other characteristics that determine whether requests are economically compatible on the same underlying hardware or can be leveraged, once isolated, for increased capital efficiency. 
Concurrent work~\cite{cascadeinfer} also partitions requests by input sequence length; 
our contribution is not limited to the use of sequence length specifically, but rather the broader formulation and production evidence for using isolation more generally as a fleet-level lever for capital efficiency.

Section \ref{sec:pillar1-non-ergodic} establishes the baseline necessity of isolation by framing unpartitioned serving as a systemic risk. 
Section \ref{sec:pillar2-cap-eff} uses production data to demonstrate how isolation directly drives down costs and surfaces new economic scaling ratios.
Section \ref{sec:pillar3-pseudo-ergodic} discusses observability and financial management, explaining how isolation fixes the broken mathematics of non-ergodic forecasting.
Section \ref{sec:pillar4-compounding-returns} explores how the foundational decision to isolate enables a secondary layer of compounding operational and hardware optimizations.
Section \ref{sec:open-challenges} discusses challenges not addressed by this paper.

\section{The End of the Ergodic Assumption: Risk and Ruin in Generative AI Infrastructure}
\label{sec:pillar1-non-ergodic}

The transition from SaaS to generative AI has fundamentally fractured the core assumptions underlying cloud serving infrastructure. 
In conventional serving, workloads exhibit predictable, bounded behavior. Compute requirements, memory footprints, and execution times are largely deterministic, allowing the infrastructure to operate under ergodic assumptions where the statistical properties of a system’s states can be reliably deduced from past averages.

Generative AI, however, represents a distinctly non-ergodic system. 
For example, input sequence lengths (and therefore resource needs) of any given request vary by orders of magnitude, and the autoregressive nature of large language models means that each inference request is an unfolding, highly path-dependent process.
Requests of varying sizes arrive at the serving infrastructure in unpredictable ways, and because generative models produce outputs one token at a time, the computational trajectory of a given prompt is acutely sensitive to its own evolving state. 
A single request might terminate after a dozen tokens or cascade into thousands, dynamically expanding its memory footprint (KV cache) at every step. Consequently, the time average of an individual request’s resource occupation diverges wildly from the ensemble average of the system.

\begin{figure}[t]
    \centering
    \includegraphics[scale=0.5]{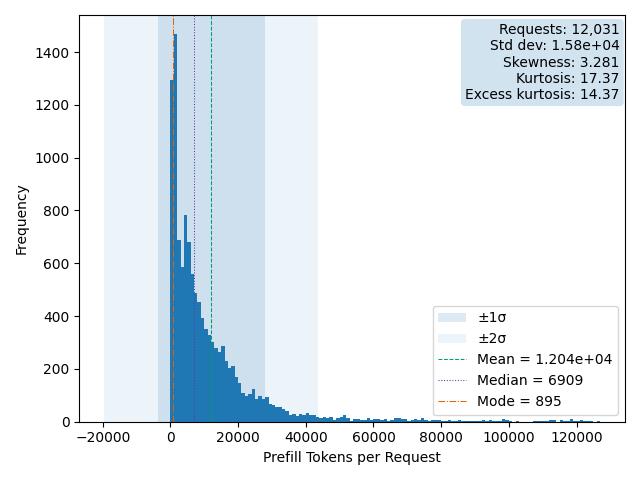}
    \vspace{-0.2cm}
    \caption{Example of non-Gaussian prompt token distributions (mooncake conversation trace) \cite{mooncake}}
    \label{fig:token-distribution-mooncake}
    \vspace{-0.8cm}
\end{figure}

Non-Gaussian distributions are often an indicator of non-ergodic behavior. 
Figure \ref{fig:token-distribution-mooncake} shows input token distributions for the mooncake conversation trace \cite{mooncake}. Similarly, in production workloads, input sequence lengths follow a highly skewed, heavy-tailed distribution. Output token lengths also appear non-Gaussian. As the field of generative AI continues to progress, the non-ergodic behavior continues to grow. Mixture-of-Expert (MoE) routing, speculative decoding, multi-token prediction, reasoning models, increased thinking depths, runtime model selection, and long-horizon agentic workloads all exacerbate non-ergodic system behavior. 
For example, while MoE architectures are typically trained with auxiliary loss functions or routing constraints that encourage an even distribution of tokens across experts~\cite{switchtransformer, gshard, expert_choice, loss_free_balancing}, real-world inference deployments can easily exhibit highly-skewed, non-Gaussian token routing distributions (e.g., repetitive content, domain shifts, or a foreign language)~\cite{openmoe, mixtral, moe_infinity}. ``Hot'' experts cause system bottlenecks and unpredictable behavior, as all-to-all communication and barrier synchronization across GPUs or clusters must wait for overloaded stragglers to finish~\cite{fastermoe, lina, janus, megascale_infer}.

This non-ergodicity poses severe challenges for conventional serving architectures, as non-ergodic systems are susceptible to \emph{ergodic ruin}~\cite{ergodicity_economics, evaluating_gambles}.
In LLM inference serving, ergodic ruin occurs when the time-average performance of a system collapses due to extreme, non-Gaussian tail events, even though the average performance looks stable and optimized on paper.
In these systems, tail events can permanently exhaust a critical resource (like GPU memory or compute budgets), halting execution or causing a system crash before the system can ``average out" its performance over time. In a capital-constrained situation, \emph{the 
increased capital requirements of the long tail can also be seen as leading to ergodic ruin}.

\begin{figure}[t]
    \centering
    \includegraphics[scale=0.5]{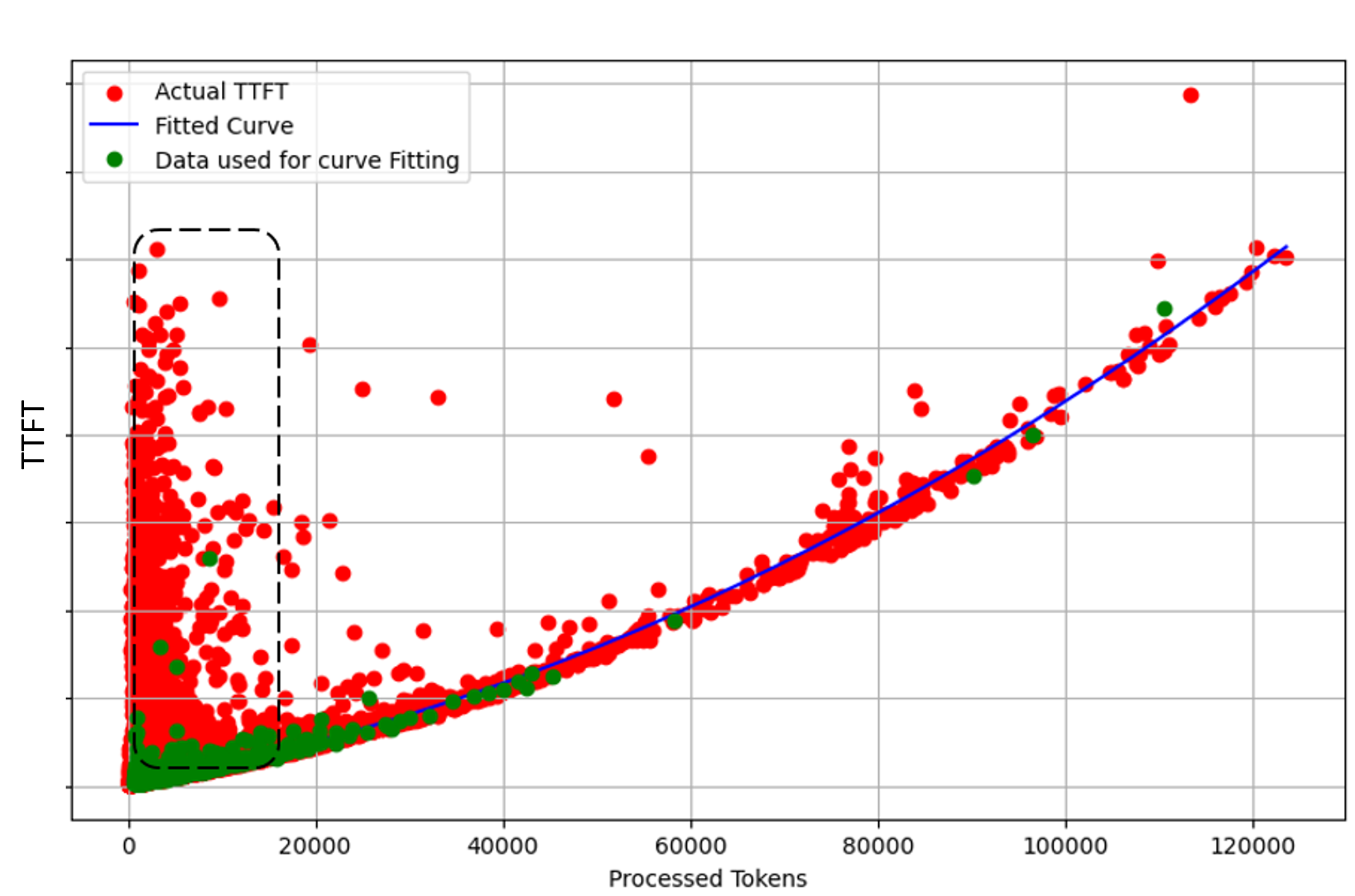} 
    \vspace{-0.2cm}
    \caption{TTFTs for production workload (peak hour) under round robin scheduling. Fitted curve estimates expected TTFT with no request interference or queuing and is hardware dependent. Factor increase over expectation: P50 = \textbf{5.9$\times$}, P75 = \textbf{21.4$\times$}, P90 = \textbf{33.4$\times$}, P95 = \textbf{52.9$\times$}, P99 = \textbf{101$\times$}}
    \label{fig:high-TTFT}
    \vspace{-0.4cm}
\end{figure}

Consider traditional load balancing algorithms, such as round robin, as an example. Round robin is highly effective at balancing load when requests are largely similar~\cite{roundrobin}. Figure \ref{fig:high-TTFT} shows what happens when round robin scheduling is applied to generative AI requests, which exhibit much more variability in their resource needs and hardware occupancy time. Small requests become blocked behind large requests, increasing TTFTs across a potentially large number of requests, resulting in an increase to aggregate latency metrics such as P95 TTFT. When P95 latencies increase, the conventional serving and provisioning playbook would introduce additional hardware to bring the latencies down, which occurs via reduced hardware utilization. In a conventional (non-AI) infrastructure serving context, P95 latencies would be a valid indicator of the need to add additional hardware, and the return on that investment would be proportionate. In generative AI serving, however, the addition of new hardware would benefit a very small number of requests (low ROIC), leave large portions of the infrastructure running at low utilization, and, worse, the buy signal would persist. As discussed in Section~\ref{sec:intro}, the imperative to keep our capital-intensive machinery running at high utilization means that \emph{we need new levers for managing performance-cost tradeoffs in a generative AI context}.  

The assumption of an ergodic underlying system permeates across the industry and academia. Simply consider the use of P95 latency metrics. P95 latencies have been used in conventional serving to represent system performance. In that context, the top 5\% of high latency requests are removed from the dataset, as they often represent some system anomaly that likely does not need to be planned for or designed around. Now consider generative AI requests and the use of P95 TTFT. In this scenario, the ommited top 5\% of the distribution does not remove anomalies but instead removes a critical part of the dataset that likely represents the requests with the highest context lengths. 

Ironically, the requests dropped from that reporting are arguably the most important to track. 
By not appropriately accounting for the costs of the requests in the fat tail, the system spreads the cost across all requests, leading to what economists would refer to as a \emph{tragedy of the commons}~\cite{tragedy_commons} in which the community of requests suffer while a small number receive a disproportionate individual benefit. 
Because heterogeneous requests share the same execution environment, severe externalities should be expected to emerge. A long-tail request steadily inflating its KV cache over thousands of forward passes can monopolize HBM and compute cycles, forcefully degrading the latency of dozens of concurrent, lightweight requests sharing the same silicon. Under peak utilization, this extreme variance inevitably leads to cascading out-of-memory (OOM) faults and prolonged ``red states," threatening systemic ruin~\cite{metastable_failures}.

\emph{Isolating request subsets can serve as a critical mechanism for systemic risk reduction}.  
By quarantining requests, e.g. based on their anticipated resource needs, infrastructure engineers can effectively contain the ``blast radius" of externalities landing on neighboring requests. Heavy, unpredictable requests are restricted to dedicated resource pools, ensuring that predictable, lightweight, high-volume tasks remain strictly insulated from resource starvation and other poor system effects.

While we can deploy more intelligent batching~\cite{qoserve, scorpio, sola, jitserve}, load balancing~\cite{llumnix, preble}, and operational guardrails, such as circuit breakers~\cite{nygard_releaseit} and admission throttling~\cite{dagor, breakwater, vtc} in an attempt to mitigate externalities, \emph{it remains broadly beneficial to explicitly break any assumptions about ergodic behavior for GenAI serving}.

This understanding encourages a departure from static, homogeneous resource allocation toward highly dynamic, system-state-aware paradigms capable of adapting to the unpredictable, path-dependent evolution of large language models. Furthermore, the underlying non-ergodic mechanics directly dictate business realities. As discussed in Section \ref{sec:pillar2-cap-eff}, the empirical cost distribution of requests does not form a predictable bell curve. Instead, it forms a highly non-Gaussian distribution characterized by severe positive skew and high kurtosis. These fat tails indicate that extreme outlier requests that consume orders of magnitude more compute and memory than the median are not rare anomalies, but a persistent, structural feature of generative AI serving that must be addressed architecturally.

\section{Capital Efficiency: Increasing Concurrency and Shaping Traffic Through Segmented Cost Structures}
\label{sec:pillar2-cap-eff}

We can use a simple isolation mechanism, segmenting traffic based on input sequence length, to demonstrate that isolation can materially improve capital efficiency when specifically directed at this objective. Even more importantly, isolation can bring \emph{cost visibility} across the range of requests. This combination surfaces \emph{new economic scaling ratios} that then enable the intentional management of the global request cost distribution towards even greater fleet-level capital efficiency. 

\textbf{Setup.} Results in this section use data gathered from one peak hour of regional production traffic spanning many instances using a co-located serving configuration across three sequence-length-delineated serving partitions consisting of roughly 70\% small, 20\% medium, and 10\% large requests. A \emph{partition} is simply defined as an execution endpoint configured to serve a subset of requests. 

\subsection{Increasing Concurrency} 
\label{sec:pillar2-cap-eff-concurrency}
Isolating requests based on input sequence length allows us to benefit from favorable system effects, reduce request interference, and drive the small sequence length partition to much higher batch sizes while staying within interactive TTFT and TBT latency constraints. 
Decode batch size is a measure of request-level parallelism in the generation phase of inference. As such, batch size is directly tied to the number of GPUs needed to serve a set of requests. Higher batch size means that the GPU cost can be amortized over the number of requests in the batch. Batch size, therefore, is a useful metric to understand capital efficiency and cost-per-request. For our production workload under interactive serving latency constraints and a co-located serving configuration deployment, the average batch size during peak hour was \textbf{14$\times$} higher in the small sequence length partition as compared to the large one. 

\begin{figure}[t]
    \centering
    \includegraphics[scale=0.4]{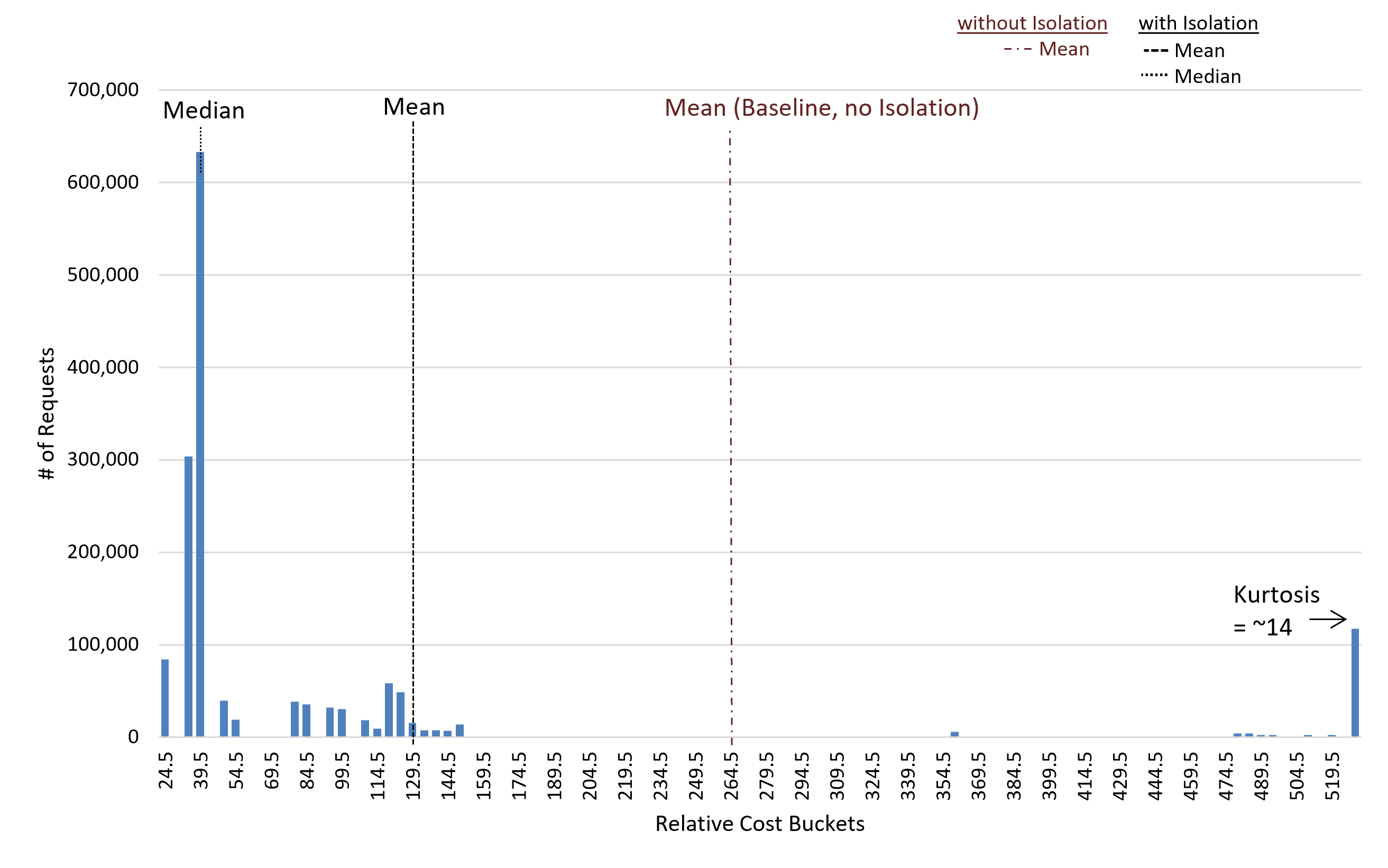}
    \vspace{-0.2cm}
    \caption{Distribution of requests by GPU cost profile for peak hour of production traffic. The mean cost achieved via isolation is \textbf{2$\times$} below the achieved mean without isolation.}
    \label{fig:cost-distr}
    \vspace{-0.5cm}
\end{figure}

In this example of isolation, grouping ``like" requests reduces interference, ensures tighter batch packing, and reduces underutilized compute, 
thereby increasing GPU utilization (and resource occupancy), increasing capital efficiency, and decreasing the cost per request. 
This isolation mechanism also has the potential to benefit from other system effects. For example, the small sequence length partition may experience more advantageous prefill caching behavior, as long requests no longer exhaust the HBM KV storage~\cite{sglang}. 
As another example, the small sequence length partition, with its increased request-level parallelism, could benefit from efficiency optimizations that leverage shared context across requests in the decode phase~\cite{flashinfer}.  

Figure ~\ref{fig:cost-distr} shows the relative cost distribution of requests 
along with the mean cost per request in the baseline without isolation. 
This simple form of isolation resulted in roughly \textbf{2$\times$} savings in the number of GPUs needed to serve the total set of requests. 
As discussed in Section~\ref{sec:intro}, high-end AI accelerators represent a massive capital expenditure for any infrastructure provider, and the return on investment is highly dependent on the utilization rates and output token production. 
By actively mitigating the destructive interference between highly variable requests and driving significantly higher batch sizes on a large percentage of requests, isolation maximizes the output token-per-second throughput per GPU. This allows the infrastructure to serve the same set of requests on a smaller hardware footprint, ultimately driving down the marginal cost of generative AI inference and ensuring that expensive compute resources are leveraged to their maximum economic potential.

\subsection{Non-Gaussian Cost Distribution}
\label{sec:pillar2-cap-eff-non-gaussian}
Viewing the distribution of requests by cost cohorts can be incredibly revealing, as it exposes a central tendency problem but also graphically exposes the non-Gaussian nature of the achievable cost distribution, thereby indicating the need to incorporate information about the higher moments as a core component of system architecture design. 
The \textbf{2$\times$} differential between the achieved mean cost per request with and without isolation
demonstrates that a `single configuration' penalty is incredibly disconnected from what is achievable with increased concurrency from targeted isolation. 
The cost achieved at a single configuration is not only unrepresentative of a majority of requests, but it can be multiples out of range with what can be achieved with isolation, thereby forcing unacceptable cost penalties (externalities) on a large percentage of requests.

While a normal cost distribution can allow for ignoring the higher statistical moments, e.g. skewness and kurtosis, the non-Gaussian nature of the achievable cost distribution indicates the need to incorporate information about the higher moments as a core component of our infrastructure design due to otherwise resulting high waste and unacceptable risk.
\emph{High skew} signals high systemic waste from targeting an average configuration, while \emph{positive kurtosis} indicates a significant unmanaged risk to profitability from presuming ergodic behavior. Simplifying assumptions that ignore these higher-order moments would be similar to investment strategies that overlook leveraged option risks, which the finance community has long since determined to be unacceptable~\cite{cvar}.
To safeguard margins and operational efficiency, core infrastructure design must incorporate these statistical realities rather than rely on previous assumptions.

\begin{figure}[t]
    \centering
    \includegraphics[scale=0.32]{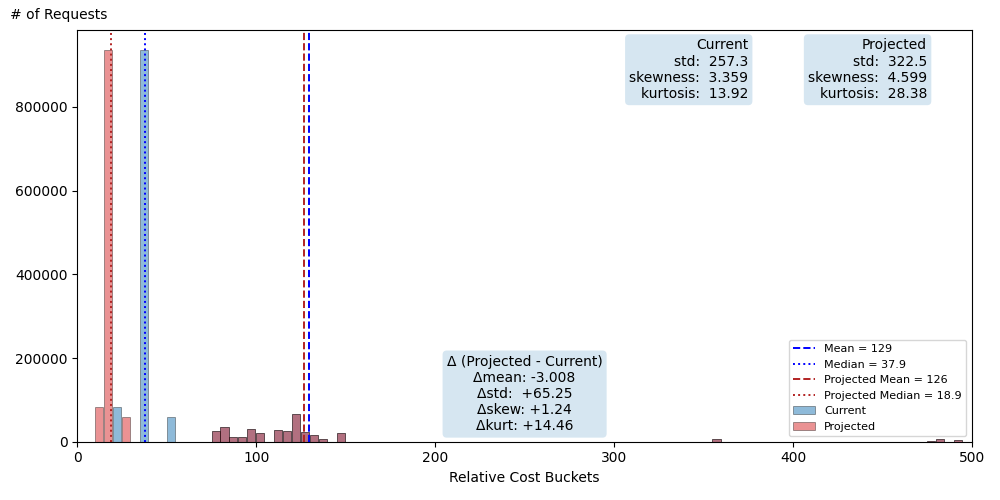}
    \vspace{-0.4cm}
    \caption{Change in summary statistics expected from the introduction of new hardware and new models that support larger maximum sequence lengths. (Zoomed in for visibility of lower moments.)}
    \label{fig:cost-distr-projected}
    \vspace{-.65cm}    
\end{figure}

\textbf{Expectations considering the evolution of hardware and models.} The non-Gaussian nature of the cost distribution 
requires dealing with the higher moments of this distribution. Looking at the expected evolution of the positive skew and excess positive kurtosis can illustrate whether these difficulties will self-resolve or become exacerbated.  
Hardware vendors and first-party silicon efforts are attempting to dramatically improve throughput-per-dollar with each subsequent hardware generation. This is mathematically equivalent to increased shifts towards lower achievable cost cohorts representing increased positive skew (3rd moment) with the median shifting more than the mean.  At the same time, model evolution towards larger models along with support for longer max sequence lengths introduces a longer tail in the distribution, thereby increasing the excess kurtosis (4th moment). As illustrated in Figure~\ref{fig:cost-distr-projected}, these two primary concerns can be expected to get more problematic, 
at least as long as the transformer architecture persists.

\subsection{New Scaling Ratios}
\label{sec:pillar2-cap-eff-ratios}
\begin{figure}[t]
    \centering
    \includegraphics[scale=0.5]{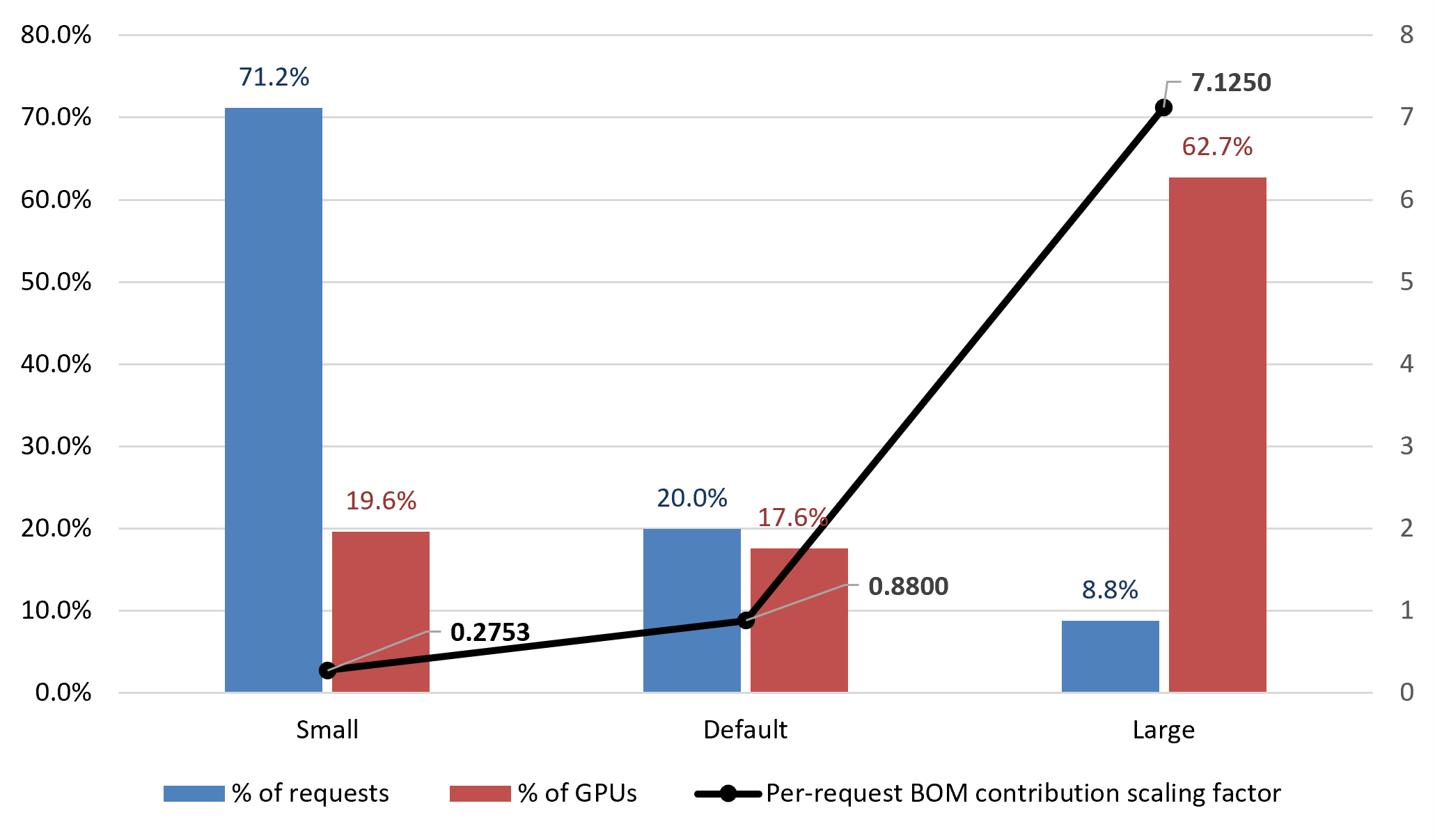}
    \vspace{-0.2cm}
    \caption{Per-partition breakdown: \% of requests and \% of GPUs; shows \textbf{25.9$\times$} cost differential before partition-specific tailoring.}
    \vspace{-.65cm}   
    \label{fig:bom}
\end{figure}

A primary benefit of isolation is the ability to bring \emph{visibility to relative costs} and \emph{enable new scaling ratios}.
When the fleet is unpartitioned, the cost of serving is obscured by a blended average leaving us blind to achievable costs.
Isolation demonstrates the opportunity to mathematically surface new scaling ratios and stark cost differentials across the fleet. 
Figure \ref{fig:bom} shows a per-partition breakdown of \% of requests and \% of GPUs, which allows us to calculate a relative cost-per-request.
The short sequence length partition handles over 70\% of requests and uses less than 20\% of the GPUs.
The long sequence length partition handles less than 9\% of requests but uses  over 60\% of the GPUs.
This places long requests as roughly \textbf{26$\times$} more costly than short ones, prior to any partition-specific optimization and tailoring. 

As the relative cost differential data demonstrates, serving highly batched, short-sequence requests yields vastly superior capital efficiency compared to long-context generation. Revealing these segmented cost structures surfaces another fleet-level capital efficiency lever enabled by the infrastructure. Specifically, it enables product and engineering teams to \emph{intentionally shape traffic, incentivizing workloads toward the segments that offer the highest capital efficiency}, thereby structurally lowering the marginal cost of inference and increasing fleet-wide capital efficiency.

\section{Pseudo-Ergodicity: Unlocking Predictability and Forecasting}
\label{sec:pillar3-pseudo-ergodic}

Isolating ``like" requests partitions the highly-variable global workload into more homogeneous, predictable subsets. By employing predictive routing or request profiling that groups requests by similar input lengths, expected output distributions, request-level tradeoff priorities, or other request-level execution characteristics, 
the serving infrastructure creates localized environments of reduced variance. 
While the global system remains fundamentally non-ergodic, these isolated pools exhibit \emph{``pseudo-ergodic"} properties. Within these segregated groups, the resource demands, execution times, and memory expansion rates are much more uniform, allowing schedulers to make reliable assumptions about batch progression. For example, infrastructure control planes can much more confidently use queue depth as a reliable input for real-time system decisions triggering dynamic request routing or proactive autoscaling, thereby driving significantly higher efficiency and performance.

\subsection{Improving the Signal-to-Noise Ratio}
Crucially, this artificially-induced predictability unlocks the ability to establish much more rigorous operational baselines, \emph{making it possible to meaningfully measure differentials from a target expectation}. In a highly non-ergodic, unisolated system, telemetry is notoriously noisy. Key health metrics like batch size, for example, will fluctuate so wildly that ``normal" is mathematically undefinable. A sudden doubling in average request latency might simply represent a temporary cluster of long prompts rather than a system failure. However, once workloads are isolated into homogeneous groups, the expected performance bounds become significantly more stable. Any deviation or differential from this established target expectation can serve as a confident signal instead of being dismissed as statistical noise. 
Because the baseline behavior within a segment is stable, observability systems can measure precise differentials from a target expectation. 
This allows observability tools to accurately detect genuine infrastructural anomalies, such as failing GPU interconnects, degraded network paths, or suboptimal routing, enabling much more precise, data-driven remediation. Remember that any degradation in generated output tokens is highly consequential to the return on the GPU investment.

\subsection{Financial Forecasting and Procurement Planning}
Traditional infrastructure planning relies on ensemble averages, an approach that mathematically fails when applied to non-ergodic generative workloads. Projecting hardware needs based on the ``average" token length of a mixed fleet results in wild inaccuracies, leading to either severe under-provisioning or wasteful over-provisioning. While mathematical averages may have some utility today, \emph{cost projections based on averages will be highly inaccurate as they do not represent a central tendency}. 

\begin{figure}[t]
    \centering
    \includegraphics[scale=0.5]{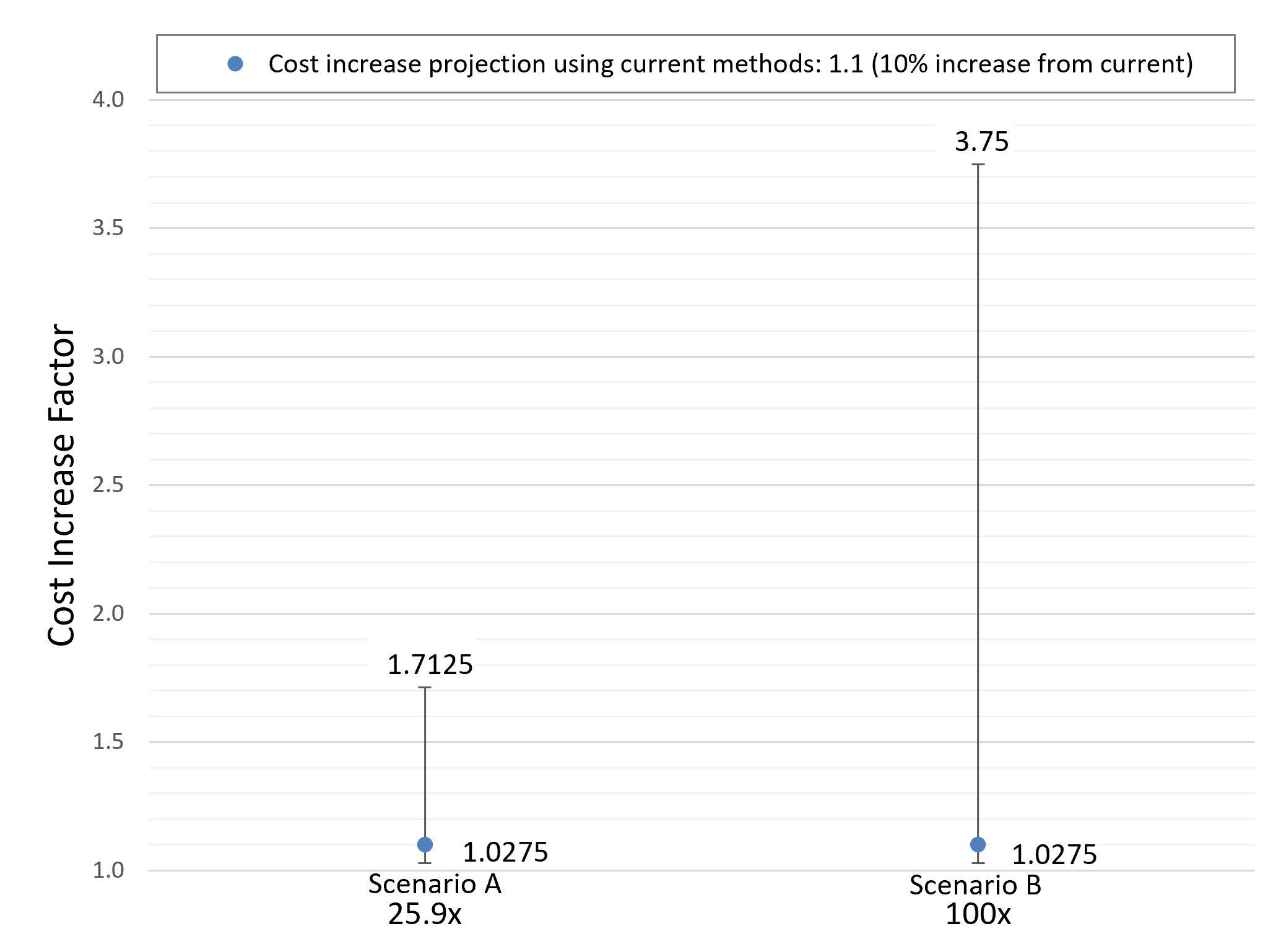}
    \caption{Range of fleet cost increase for 10\% increase in traffic for two scenarios of relative cost differential between partitions. Top bar indicates fleet cost increase if all growth occurs in the large sequence length partition. Lower bar indicates the cost increase if all growth occurs in the small sequence length partition.}
    \vspace{-.65cm}   
    \label{fig:cost-projection-2-scenarios}
\end{figure}

Figure~\ref{fig:cost-projection-2-scenarios} illustrates how projections based on a mean can be wildly inaccurate, depending on which type of traffic grows. Consider an example where we assume the fleet is currently at capacity, and we anticipate a 10\% increase in traffic. The status quo projection would plan for a 10\% increase in fleet costs and hardware needs. Now consider the scenario described in Section \ref{sec:pillar2-cap-eff} where the relative cost differential between the small sequence length and large sequence length partitions is \textbf{25.9$\times$}. If all of the traffic growth occurred within the small partition, this would only increase fleet costs by 2.75\% rather than the projected 10\%. More worrisome is if the growth occurs in long sequence length group. In that case, fleet costs and hardware needs would grow by 71.52\% rather than the projected 10\%.

Figure \ref{fig:cost-projection-2-scenarios} also illustrates how this issue is exacerbated as the relative cost differential between requests grows, which is the trend we expect moving into the future. In a scenario where the cost differential spans two orders of magnitude, a 10\% increase in the costly (low capital efficiency) partition would require a 275\% increase in the GPU fleet. That is, the fleet size would need to grow by an additional \textbf{2.75$\times$} to support this 10\% growth. 

Isolating to restore pseudo-ergodic, localized predictability not only improves observability and an understanding of performance-to-target, but also restores the validity of cost projections, allowing financial models to accurately forecast hardware and infrastructure scaling requirements. When variance is mathematically constrained, schedulers can much more accurately predict memory allocation rates; auto-scaling mechanisms can provision new nodes based on more reliable, smoothed throughput trajectories rather than erratic, momentary latency spikes; and infrastructure operations and planning can shift from reactive firefighting to deterministic planning. 

\section{Compounding Returns: Fleet Heterogeneity, Targeted Optimization, and Hardware-Software Co-Design}
\label{sec:pillar4-compounding-returns}

The capital efficiency gained through workload isolation is not a one-time, static gain; \emph{it is a foundational platform lever that unlocks a secondary layer of compounding returns}.

\subsection{Innovation and Targeted Optimizations}

In an unpartitioned, mixed-serving environment, software engineering and the integration of research outcomes are frequently paralyzed by the fear of collateral regressions. A kernel or scheduling optimization can easily be beneficial in some cases and detrimental in others. In the sequence length isolation example, some optimizations that dramatically benefit long context can easily add disproportional overheads or unnecessary tradeoffs (e.g. approximation) to requests with small sequence lengths. 
When all traffic shares the same execution endpoints, a new optimization must essentially improve the blended average across all sequence lengths to justify deployment. Consequently, high-value optimizations that only benefit a specific subset of the workload are regularly shelved.

Isolation breaks down this barrier to innovation and muted capital efficiency improvements. By partitioning the fleet into homogeneous, pseudo-ergodic segments, engineering and research teams can tailor the serving stack choices, hardware backing, provisioning, dynamic allocation strategies, etc. to the exact mechanical bottlenecks of each subset. The short-sequence set can be aggressively optimized for minimal scheduling overhead and high-concurrency batching, while the long-sequence set can deploy heavy memory-management schemes and speculative decoding strategies without risking performance degradation for the rest of the fleet. 

This tailored approach extends directly to Service Level Agreements (SLAs) and provisioning strategies. Rather than forcing the entire infrastructure to over-provision against a single, rigid latency target, isolation allows performance evaluation based on a differential to partition-specific targets. For example, we can chose to provision for stricter TTFT targets for short sequence length requests while strategically relaxing TTFT expectations for long sequence length requests or reasoning requests with deep thinking.

\subsection{Clarity on Investment Decisions and Heterogeneous Hardware Deployment}

While Section \ref{sec:pillar2-cap-eff} discussed the benefit of bringing visibility to relative cost differentials, many critical questions remain beyond 
``How much do things cost?"
For example: How should we deploy the hardware we already have? How do we do so in the most cost-effective way? When should we invest in new hardware? And what would the return on that capital look like? 

Isolation brings unprecedented clarity to these questions including the expected Return on Invested Capital (ROIC).
To start, surfacing relative cost differentials (Section~\ref{sec:pillar2-cap-eff}) combined with more accurately predicting expected throughput (Section~\ref{sec:pillar3-pseudo-ergodic}) \emph{allows for fine-grained investment decisions regarding the deployment of new hardware}. For comparing potential investments, ROIC can be viewed in terms of output tokens per second per investment unit, i.e. output tokens per second per hardware unit. 
While the average ROIC across the fleet or across the global set of requests may not exceed a predefined hurdle rate, a more fine-grain breakdown of investment options allows capital-efficient partitions, like the small sequence length partition in the example use case, to receive resources rather than starve. Mathematically, this is a high return investment of capital, which, as mentioned in Section~\ref{sec:pillar2-cap-eff-ratios}, allows for \textbf{26$\times$} growth per investment unit as compared to alternative investment options. 

Further, this granular ROIC visibility solves one of the most pressing challenges in modern AI infrastructure, which is how to most cost effectively deploy a heterogeneous hardware fleet. Given the rapid cadence of model and hardware generations, it is safe to assume that a production infrastructure will not operate on 100\% state-of-the-art (SOTA) silicon; fleets are inevitably a mix of generational architectures, which themselves represent a basket of tradeoffs, e.g. varying model architectures, compute-to-memory ratios, cost points, etc. In a mixed-request setup, we can make some gross decisions based on the model, for example, or prefill versus decode server hardware in a disaggregated serving configuration \cite{splitwise}. But, at a finer granularity, it remains largely unclear how to deploy heterogeneous hardware cost effectively while avoiding high variability and sometimes unacceptable performance degradations. In an isolated serving architecture, however, \emph{workload subsets can be precisely matched to the physical reality of the silicon}. Isolation can, therefore, preserve the scarcest SOTA accelerators, expensive and/or supply-chain constrained networking components (e.g. Infiniband), and maximum-bandwidth HBM exclusively for the requests that need it, such as long-context and deep-reasoning segments that genuinely require massive memory bandwidth to avoid stalling during decode. Notably, this \emph{isolation also acts as a viable approach to extending a GPU's useful life}, which is critical for maximizing its lifetime output token production.

In addition to informing hardware deployment and investment decisions, \emph{isolation can additionally inform on the best return on engineering or research investments}. For example, rather than using the conventional lever of increasing hardware provisioning to improve performance on a capital inefficient partition, such as the long sequence length partition, isolation surfaces a clear opportunity to target research resources at optimizations that improve capital efficiency, creating space for growth without adding more hardware. Such a situation also provides clear, measurable value add for any results of those research efforts. 

\begin{figure}[t]
    \centering
    \includegraphics[scale=0.45]{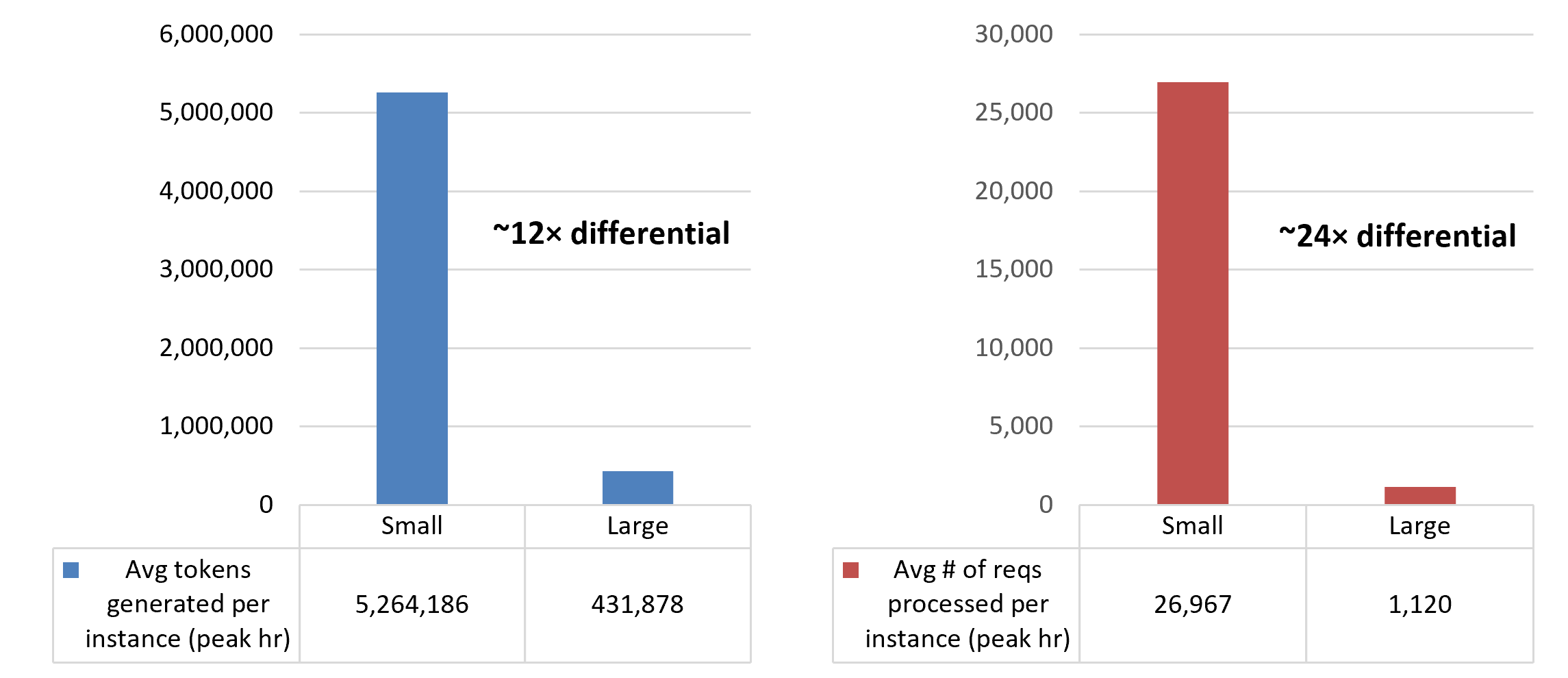}
    \vspace{-0.2cm}
    \caption{Comparing gains per-dollar-spend via proxies for utility produced (output tokens) and users impacted (requests processed).}
    \label{fig:proxies}
\end{figure}

We can also consider gains per-dollar-spend via proxies for utility produced and users impacted. 
In generative AI, we build utility from the output tokens of the requests. Thus, \emph{output tokens} can act as a useful proxy for \emph{utility produced} for a given investment. Increased output tokens per second per GPU generally would indicate greater achieved utility for a given GPU investment (increased ROIC). Similarly, \emph{requests processed} can act as a proxy for the \emph{number of users affected} per dollar spend, where more requests processed per time window indicates a larger number of users affected.
Figure \ref{fig:proxies} shows that, in the sequence-length-based isolation use case investigated, the short sequence length partition processes the overwhelming majority of distinct requests. Investing in optimizations or capacity for this segment yields the highest return in terms of maximizing the number of users affected per dollar of capital deployed, as well as the utility produced per dollar spend. During peak hour, the short sequence length partition generates \textbf{12$\times$} more tokens per hardware instance on average and processes \textbf{24$\times$} more requests per instance on average as compared to the large sequence length partition.

\subsection{Informing Next-Generation Silicon Design}
Viewed from a macro-economic perspective, \emph{one of the most significant benefits of workload isolation is its ability to inform the multi-year trajectory of custom silicon co-design}. Designing an ASIC or custom accelerator for an unpartitioned, non-ergodic workload forces hardware architects into expensive, general-purpose compromises. This typically results in over-provisioning FLOPs, HBM, or both on every single die to survive worst-case tail variance. By isolating workloads and stabilizing their statistical profiles, we unmask the distinct physical bottlenecks of each segment. Infrastructure telemetry transitions from noisy, blended averages into a much more precise specification sheet for next-generation hardware needs. Armed with this data, silicon teams can move away from monolithic, catch-all architectures and intentionally design specialized classes of silicon 
engineered specifically to address their needs where third party vendors are otherwise falling short.

\subsection{Recap of Compounding Returns}

To recap, mixed-request serving can suffer from:
\vspace{-0.15cm}
\begin{itemize}[nosep]
    \item Aggregate metrics and an aggregate buy signal
    \item ROIC that is unclear and ultimately much lower than what is achievable with isolation
    \item An aggregate ROIC that will not perform well in a competition for supply-constrained resources (SOTA hardware) in comparison to alternatives
    \item Limited clarity on the most cost-advantageous ways of deploying new SOTA hardware and existing heterogeneous hardware
    \item Limited clarity on alternative levers that could be pulled, e.g. research investment as an alternative to increased hardware provisioning
    \item Limited visibility to measurable returns on research investment and dampened innovation due to a one-size-fits-all, `no regression' bar
    \item Few structures for incentivizing behavioral changes that result in lower cost requests
    \item Inaccurate cost projections and capacity forecasting, as well as suboptimal subsequent pricing strategies due to central tendency problems    
\end{itemize}

Isolation addresses all of the above by surfacing: 
\vspace{-0.15cm}
\begin{itemize}[nosep]
    \item Per-partition relative cost visibility
    \item Per-partition tradeoff objectives
    \item Per-partition metrics
    \item Per-partition optimization (tailoring) that further improves capital efficiency
    \item Per-partition hardware allocation strategies
    \item Per-partition demand signals
    \item Per-partition ROIC calculations
    \item Per-partition provisioning strategies
    \item Per-partition growth projections
\end{itemize}

\section{Tradeoffs and Open Challenges}
\label{sec:open-challenges}

While the empirical findings in this paper use input sequence length to demonstrate the benefits of isolation on capital efficiency, this specific isolation mechanism is only intended to represent a sufficient example proof-of-concept for targeting capital efficiency. Input sequence length serves as an easy-to-understand, simple pivot point that is largely deterministic (number of bytes acts as a sufficiently accurate proxy) and known a priori. However, the overarching, intellectual contribution of this work is \emph{the establishment of isolation as a macro, fleet-level lever for capital efficiency}. Sequence length is merely one dimension along which a non-ergodic system can be partitioned.

\subsection{Tradeoffs}
Note that motivating workload isolation is not a call to fracture the infrastructure into an infinite number of hyper-specialized micro-partitions. There are important tradeoffs to consider including engineering effort, maintenance burden, prefill cache hit rate and others.
While ``fleet fragmentation" and ``stranded capacity" may at first seem to be an inevitable outcome of isolation, the extent to which this actually occurs is dependent on other system architecture choices and their alignment with (or divergence from) some conventional serving assumptions. We leave a detailed discussion of these choices as outside of the scope of this paper. 
It is true that having a sufficient concentration of homogeneous requests to keep the silicon saturated is an important consideration. 
If a partition is defined too narrowly and/or the serving infrastructure includes architectural barriers that limit provisioning flexibility, the realized capital efficiency benefits can be reduced by underutilized capacity waiting for enough ``like" requests to arrive. Again, the extent to which this actually occurs is dependent on other architectural choices and scale.

As usual, there is a tradeoff between generality (widespread applicability) and the efficiency benefits that come from special-purpose design decisions. This debate typically occurs in the context of special-purpose silicon design; however, it is useful to note that in a generative AI inference serving stack, we have many software-based configuration options at our disposal. Examples include chunk size, parallelization scheme combinations, KV cache eviction policies, load balancing algorithms, etc. We believe that the benefits gained from targeting capital efficiency improvements will increasingly justify system architectures that expose an increasing number of configuration knobs that materially affect cost, accuracy, and performance. The expected direction of travel can be characterized as: 
\vspace{-0.2cm}
\begin{itemize}[nosep]
    \item an increasing number of exposed knobs
    \item with increasingly fine-grain control
    \item across an increasing portion of the serving stack
    \item that are increasingly available for execution time decisions, as opposed to design time ones  
\end{itemize}
\vspace{-0.1cm}
We leave an in-depth discussion on tailoring for future work. 

It is also beneficial to revisit tradeoff assumptions concerning increased complexity and maintenance burdens and the realized capital efficiency benefits. 
In reasoning about this tradeoff, it is worth noting a material change from a conventional context. In conventional (non-AI) serving, the majority of hyperscaler costs were concentrated in engineering. In contrast, as discussed in Section~\ref{sec:intro}, generative AI is highly capital intensive and the hardware and data center capital costs now constitute the primary pain point relative to costs. The net result is that, as scale and total capital requirements increase, \emph{we must challenge previously held heuristics regarding 
expectations on the relative cost-benefit analysis between additional operational burdens and reduced capital requirements}.

\subsection{Optimization across Isolation Mechanisms and Boundary Selections}
Determining where and how to draw these isolation boundaries exposes a complex, multi-dimensional optimization problem that this paper intentionally leaves open for future exploration. Because request homogeneity can be defined across numerous axes, infrastructure architects must navigate a vast state space of potential isolation criteria, each carrying distinct tradeoffs, opportunities, and challenges.

\textbf{Ex-Ante vs. Ex-Post Observability.} While input sequence length is known immediately, output generation length is governed by the stochastic, path-dependent evolution of the model. The workload investigated had strong correlations between input and output sequence lengths, but a small input sequence length request can still trigger an extremely long response or reasoning chain, which can be difficult in some cases to know ahead of time. Therefore, isolating along some dimensions might require equipping the platform with lightweight predictive classifiers or speculative length-estimation models. In many cases, more accurate predictions would come at the expense of latency or cost overheads, while trading accuracy can increase the risk of misrouting and/or the need for mid-generation migration protocols when a request breaches its expectation, e.g. memory requirements.

\textbf{Information Passing.} Besides lightweight predictive classifiers or learned estimation techniques, there is an opportunity for the serving infrastructure to support \emph{the passing of request taxonomy information, metadata, and other hints} that communicate \emph{isolation-relevant criteria}. This approach effectively \emph{`shapes'} a request for more intelligent, capital-efficient routing. This efficiency-relevant information might be learned offline, e.g. using spare compute cycles to analyze a previous conversation turn or agent activity, or it could be explicitly communicated as part of the API interface. 

\textbf{``Likeness", Cost Levels, and Economic Value.} As frontier models diversify into Mixture of Experts (MoE) architectures, multi-modal encoders, and dynamic LoRA adapter topologies, \emph{``likeness"} clearly extends beyond token counts. Two requests of identical sequence length may exhibit completely different execution profiles depending on expert routings, the shape of active parameters, multi-modal needs, etc. 
Requests also vary wildly in their temporal urgency, as well as in their expected utility. For example, synchronous, user-facing conversational turns might require strict TTFT bounds, whereas asynchronous background agents can tolerate substantial queuing delays in exchange for maximum batch density.
 
 As we look into the future, there are many request characteristics and tradeoff priorities beyond latency constraints that will contribute to costs and which would only add value to a subset of global requests. One simple example is batch invariant kernels, which can provide determinism (repeatable LLM outputs), though at a 20-30\% performance hit~\cite{thinkingmachines_determinism}. While deterministic behavior may be of significant value to some requests, it would be suboptimal to carry this cost for the requests where this behavior does not add value. 
 Choices for handling AI Memory, post-processing for security concerns, and AI responsibility-based enhancements represent a few other examples. 

\section{Conclusion}
\label{sec:conclusion}

The economic context of capital-intensive generative AI serving requires the consideration of inconvenient tradeoff methodologies to improve capital efficiency, including the use of isolation.
Not only can isolation act as a fleet-level lever for increasing capital efficiency, it additionally lays the architectural groundwork that can serve as a launching point for compounding gains in capital efficiency. This includes surfacing \emph{new scaling ratios} that serve as traffic magnets, as well as near-infinite opportunities for \emph{narrow-band tailoring}. 

Solving for the optimal isolation mechanisms, boundaries, and number of partitions across a continuously changing, multidimensional space
is a complex, open challenge with the definition of ``optimal" drifting as the cost of capital increases. However, even without a fixed definition of optimal, our exploration has demonstrated that isolation can provide `multiples' in capital efficiency savings, and we have only scratched the surface of what is possible. 
As the parameters of this optimization space evolve, fleet-wide capital efficiency provides an emerging primary compass for navigating architectural tradeoffs and guiding the underlying system design decisions.


\section*{Acknowledgments} 
Special thanks to Chris Mitchell and Alan Tus, whose close collaboration and operational partnership were instrumental in taking this work to live production serving. Their commitment to improving GPU efficiency made these production evaluations possible, and we look forward to deepening our collaboration through end-to-end co-design on future systems initiatives.

\bibliography{references}
\bibliographystyle{mlsys2025}

\section*{AI Usage Disclosure} AI was \emph{not} used for the intellectual contributions, ideation, experiment design, implementation, or data generation included in this report.
AI \emph{was} used to polish and shorten certain sections during the editing process to meet submission requirements.




\end{document}